\documentclass[runningheads]{llncs}%
\usepackage{amsmath}
\usepackage{amssymb}
\usepackage{amsfonts}
\usepackage{graphicx}%
\providecommand{\U}[1]{\protect\rule{.1in}{.1in}}
\titlerunning{Assimilation of fire perimeters and satellite detections}
\institute{
  HPCA4SE research group, Computer Architecture and Operating Systems Department,\\
  Universitat Aut{\`o}noma de Barcelona\\
  08193 Bellaterra, Spain\\
  \email{\{angel.farguell,ana.cortes\}@uab.cat}
\and
Department of Mathematical and Statistical Sciences, \\
University of Colorado Denver, 1201 Larimer St. \\
Denver, CO 80204, United States of America\\
\email{\{angel.farguellcaus,james.haley,jan.mandel\}@ucdenver.edu}
\and
Department of Atmospheric Sciences \\
University of Utah, 135 S 1460 East Rm 819 (WBB) \\
Salt Lake City, Ut 84112-0110, United States of America\\
\email{adam.kochanski@utah.edu}
}
\authorrunning{A.~Farguell Caus, J.~Haley, A.K.~Kochanski, A.~Cort{\'e}s Fit{\'e}, J.~Mandel}
\begin{document}

\title{Assimilation of fire perimeters and satellite detections by minimization of
the residual in a~fire spread model}
\author{Angel Farguell Caus\inst{1}\inst{2}
\and James Haley\inst{2}
\and Adam K. Kochanski\inst{3}\\Ana Cort\'{e}s Fit{\'e}\inst{1}
\and Jan Mandel\inst{2}}
\maketitle

\begin{abstract}
Assimilation of data into a fire-spread model is formulated as an optimization
problem. The level set equation, which relates the fire arrival time and the
rate of spread, is allowed to be satisfied only approximately, and we minimize
a norm of the residual. Previous methods based on modification of the fire
arrival time either used an additive correction to the fire arrival time, or
made a position correction. Unlike additive fire arrival time corrections, the
new method respects the dependence of the fire rate of spread on diurnal
changes of fuel moisture and on weather changes, and, unlike position
corrections, it respects the dependence of the fire spread on fuels and
terrain as well. The method is used to interpolate the fire arrival time
between two perimeters by imposing the fire arrival time at the perimeters as
constraints.

\end{abstract}



%



\section{Introduction}

Every year, millions of hectares of forest are devastated by wildfires. This
fact causes dramatic damage to innumerable factors as economy, ecosystem,
energy, agriculture, biodiversity, etc. It has been recognized that the recent
increase in the fire severity is associated with the strict fire suppression
policy, that over last decades has led to significant accumulation of the
fuel, which when ignited makes fires difficult to control. In order to reverse
this effect, prescribed burns are routinely used as a method of fuel reduction
and habitat maintenance~\cite{Outcalt-2004-FMR,Stephens-2005-FFP}. The
previous strategy of putting out all wildland fires is becoming replaced by a
new approach where the fire is considered as a tool in the land management
practice, and some of the fires are allowed to burn under appropriate
conditions in order to reduce the fuel load and meet the forest management goals.

Fire management decisions regarding both prescribed burns, as well as wildland
fires, are very difficult. They require a careful consideration of potential
fire effects under changing weather conditions, values at risk, firefighter
safety and air quality impacts of wildfire smoke~\cite{Yoder-2004-LIP}. In
order to help in the fire management practice, a wide range of models and
tools has been developed. The typical operational models are generally
uncoupled. In these models, elevation data (slope) and fuel characteristics
are used together with ambient weather conditions or general weather forecast
as input to the rate of spread model, which computes the fire propagation
neglecting the impact of the fire itself on local weather conditions (see
BehavePlus~\cite{Andrews-2007-BFM}, FARSITE~\cite{Finney-1998-FFA} or
PROMETHEUS~\cite{Tymstra-2010-DSP-x}). As computational capabilities increase,
a new generation of coupled fire-atmosphere models become available for fire
managers as management tools. In a coupled fire-atmosphere model, weather
conditions are computed in-line with the fire propagation. This means that the
state of the atmosphere is modified by the fire so that the fire spread model
is driven by the local micrometeorology modified by the fire-released heat and
moisture fluxes. CAWFE~\cite{Coen-2013-MWF}, WRF-SFIRE~\cite{Mandel-2011-CAF},
and FOREFIRE/Meso-NH~\cite{Filippi-2011-SCF}, are examples of such models,
coupling CFD-type weather models with semi-empirical fire spread models. This
approach is fundamentally similar to so-called physics-based models like
FIRETEC~\cite{Linn-2002-SWB} and WFDS~\cite{Mell-2007-PAM}, which also use CFD
approach to compute the flow near the fire, but focus on flame-scale processes
in order to directly resolve combustion, and heat transfer within the fuel and
between the fire and the atmosphere. As the computational cost of running
these models is too high to facilitate their use as forecasting tools, this
paper focuses on the aforementioned hybrid approach, where the fire and the
atmosphere evolve simultaneously affecting each other, but the fire spread is
parameterized as a function of the wind speed and fuel properties, rather than
resolved based on the detailed energy balance.

\begin{figure}[ptb]
\centering
\includegraphics[scale=0.4]{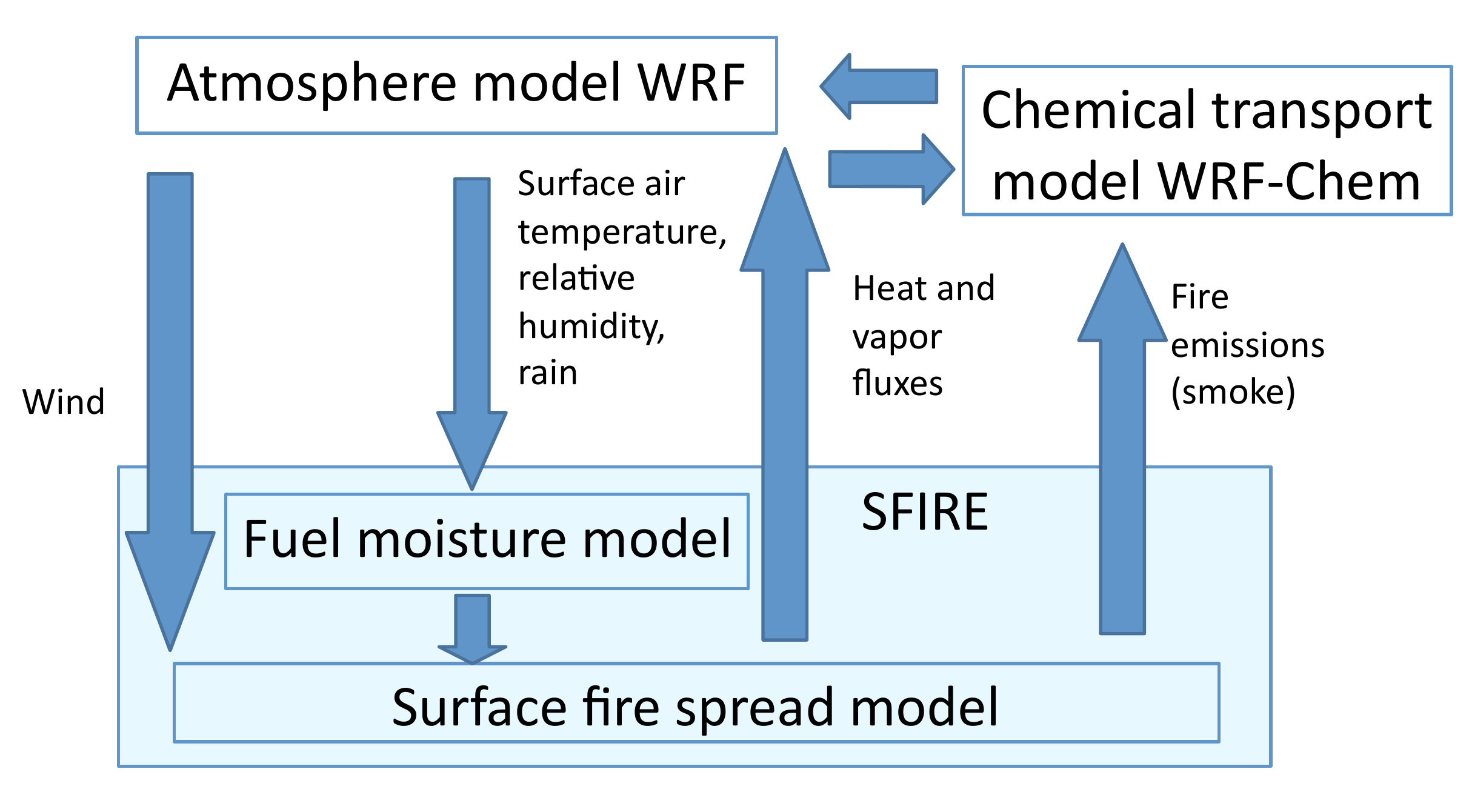}\caption{Diagram of the
model coupling in WRF-SFIRE}%
\label{fig:diagram}%
\end{figure}

This article describes upcoming data assimilation components for
the coupled fire-atmosphere model WRF-SFIRE~\cite{Kochanski-2016-TIS,Mandel-2014-RAA},
 which combines a mesoscale numerical weather prediction system, WRF~\cite{Skamarock-2008-DAR}, 
with a surface fire behavior model implemented by a level set method, a fuel moisture model~\cite{Vejmelka-2016-DAD},
and chemical transport of emissions. The coupling between the models is
graphically represented in the diagram in Fig.~\ref{fig:diagram}. The fire
heat flux modifies the atmospheric state (including local winds), which in
turn affects fire progression and the fire heat release. WRF-SFIRE has evolved from CAWFE
\cite{Clark-2004-DCA,Coen-2005-SBE}. An~earlier version~\cite{Mandel-2011-CAF} 
is distributed with the WRF\ release as WRF-Fire~\cite{Coen-2013-WCW}, and it
was recently improved by including a~high-order accurate level-set method~\cite{Munoz-2018-AFS}.

The coupling between fire and atmosphere makes initialization of a fire from satellite detections and/or
fire perimeters particularly challenging. In a coupled numerical
fire-atmosphere model, the ignition procedure itself affects the atmospheric
state (especially local updrafts near the fire line and the near fire winds).
Therefore, particular attention is needed during the assimilation process in
order to assure that realistic fire-induced atmospheric circulation is
established at the time of data assimilation. One possible solution to this
problem, assuring consistency between the fire and the atmospheric models, is
defining an artificial fire progression history, and using it to replay the
fire progression prior to the assimilation time. In this case, the heat
release computed from the synthetic fire history is used to spin up the
atmospheric model and assure consistency between the assimilated fire and the
local micro-meteorology generated by the fire itself.

Fire behavior models run on a mesh given by fuel data availability, typically
with about 30m resolution and aligned with geographic coordinates. The mesh
resolution of satellite-based sensors, such as MODIS and VIIRS, however, is
typically 375m-1.1km in flight-aligned swaths. These sensors provide
planet-wide coverage of fire detection several times daily, but data may be
missing for various reasons and no detection is possible under clouds; such
missing pixels in the swath are marked as not available or as a cloud, and
distinct from detections of the surface without fire. Because of the missing
data, the statistical uncertainty of detections, the uncertainty in the actual
locations of active fire pixels, and the mismatch of scales between the fire
model and the satellite sensor, direct initialization of the model from
satellite fire detection polygons~\cite{Coen-2013-USR} is of limited value at
the fuel map scale. Therefore, the satellite data should be used to steer such
models in a statistical sense only.

In this study, we propose a new method of fitting fire arrival time to data,
which can be used to generate artificial fire history, which can be used to
spin up the atmospheric model for the purpose of starting a simulation from a
fire perimeter. In combination with detection data likelihood, the new method
can be used also to assimilate satellite fire detection data. This new method,
unlike position or additive time corrections, respects the dependence of the
fire rate of spread on topography, diurnal changes of fuel moisture, winds, as
well as spatial fuel heterogeneity.

\section{Fire spread model}

The state of the fire spread model is the fire arrival time $T\left(
x,y\right)  $ at locations $\left(  x,y\right)  $ in a rectangular simulation
domain $\Omega\subset\mathbb{R}^{2}$. The isoline $T\left(  x,y\right)  =c$ is
then the fire perimeter at time $c$. The normal vector to the isoline is
$\nabla T/\left\Vert \nabla T\right\Vert $. The rate of spread in the normal
direction and the fire arrival time at a location on the isoline then satisfy
the eikonal equation%
\begin{equation}
\left\Vert \nabla T\right\Vert =\frac{1}{R}.\label{eq:eikonal}%
\end{equation}
We assume that $R$ depends on location (because of different fuel, fuel
moisture, and terrain) and time (because of wind and fuel moisture changing
with time). Rothermel's model~\cite{Rothermel-1972-MMP-x} for 1D\ fire spread
postulates%
\begin{equation}
R=R_{0}(1+\phi_{w}+\phi_{s}),\label{eq:roth}%
\end{equation}
where $R_{0}$ is the omnidirectional rate of spread, $\phi_{w}$, the wind
factor, is a function of wind in the spread direction, and $\phi_{s}$, the
slope factor, is a function of the terrain slope. The 1D model was adapted to
the spread over 2D landscape by postulating that the wind factor and the slope
factor are functions of the components of the wind vector and the terrain
gradient in the normal direction. Thus,%
\begin{equation}
R=R\left(  x,y,T\left(  x,y\right)  ,\nabla T\left(  x,y\right)  \right)
.\label{eq:R}%
\end{equation}

The fire spread model is coupled to an atmospheric model. The fire emits
sensible and latent heat fluxes, which change the state of the atmosphere, and
the changing atmospheric conditions in turn impact the fire
(Fig.~\ref{fig:diagram}). Wind affects the fire directly by the wind factor,
and temperature, relative humidity and rain affect the fire through changing
fuel moisture.

The fire model is implemented on a rectangular mesh by finite differences. For
numerical reasons, the gradient in the eikonal equation (\ref{eq:eikonal})
needs to be implemented by an upwinding-type method~\cite{Osher-2003-LSM}, which avoids instabilities
caused by breaking causality in fire propagation: for the computation of
$\nabla T$ at a location $\left(  x,y\right)  $, only the values from the
directions that the fire is coming from should be used, so the methods switch
between one-sided differences depending on how the solution evolves. Sophisticated methods
of upwinding type, such as ENO or flux-limiters~\cite{Rehm-2009-FPU}, aim to use more accurate
central differences and switch to more stable one-sided upwind differences
only as needed. Unfortunately, the switching causes the numerical gradient of
$T$ at a mesh node become a nondifferentiable function of the values of $T$ at
that point and its neighbors. In addition, we have added a penalty term to
prevent the creation of local minima. It was observed
in~\cite{Mandel-2009-DAW} that if, in the level set method, a local minimum
appears on the boundary, its value keeps decreasing out of control; we have
later found out that this can in fact happen anywhere in the presence of
spatially highly variable rate of spread, and we have observed a similar
effect here during the minimization process.

\section{Fitting the fire spread model to data}

\subsection{Minimal residual formulation}

\label{sec:eikonal}

Consider the situation when the two observed fire perimeters $\Gamma_{1}$ and
$\Gamma_{2}$ at times $T_{1}<T_{2}$ are known, and we are interested in the
fire progression between the two perimeters. Aside from immediate uses
(visualization without jumps, post-fire analysis), such interpolation is
useful to start the fire simulation from the larger perimeter $\Gamma_{2}$ at
time $T_{2}$ by a spin-up of the atmospheric model by the heat fluxes from the
interpolated fire arrival time between the fire perimeters; the coupled model
can then start from perimeter $\Gamma_{2}$ at time $T_{2}$ in a consistent
state between the fire and the atmosphere. Interpolation between an ignition
point and a perimeter can be handled the same way, with the perimeter
$\Gamma_{1}$ consisting of just a single point.

In this situation, we solve the eikonal equation (\ref{eq:eikonal}) only
approximately,
\begin{equation}
\left\Vert \nabla T\right\Vert \approx\frac{1}{R} \label{eq:approx}%
\end{equation}
imposing the given fire perimeters as constraints,%
\begin{equation}
T=T_{1}\text{ at }\Gamma_{1}\text{,}\quad T=T_{2}\text{ at }\Gamma_{2}.
\label{eq:perimeter-conditions}%
\end{equation}
We formalize (\ref{eq:approx}) as the minimization problem
\begin{equation}
J(T)=\left(  \int_{\Omega}\left\vert f(\left\Vert \nabla T\right\Vert _{2}%
^{2},R^{2})\right\vert ^{p}\right)  ^{1/p}\rightarrow\min_{T}\text{ subject to
(\ref{eq:perimeter-conditions}),} \label{eq:constrained-min}%
\end{equation}
where $f\left(  x,y\right)  $ is a function such that $f\left(  x,y\right)
=0$ if and only if $xy=1$, and $\Omega$ is the simulation domain. We mostly
use the function $f\left(  x,y\right)  =1-xy$ but other functions, such as
$f\left(  x,y\right)  =x-1/y$ have advantages in some situations. There are no
boundary conditions imposed on the boundary of $\Omega$.

\subsection{Discretization and the constraint matrix}

\label{sec:H} The fire simulation domain is discretized by a logically
rectangular grid (aligned approximately with longitude and latitude) and
perimeters are given as shape files, i.e., collections of points on the
perimeter. We express (\ref{eq:perimeter-conditions}) in the form
\begin{equation}
HT=g, \label{eq:perimeter-cond-discrete}%
\end{equation}
where $H$ is a sparse matrix. Since the points in the shape files do not need
to lie on the grid, the rows of $H$ are the coefficients of an interpolation
from the grid to the points in the shape files, which define the perimeters.
We find the coefficients from barycentric interpolation. The rectangles of the
grid are split into two triangles each, and, for each triangle, we compute the
barycentric coordinates of the points in the shapefile, i.e., the coefficients
of the unique linear combination of the vertices of the triangle that equals
to the point in the shape file. If all 3 barycentric coordinates are in
$\left[  0,1\right]  $, we conclude that the point is contained in the
triangle, the barycentric coordinates are the sought interpolation
coefficients, and they form one row of $H$. For efficiency, most points in the
shapefile are excluded up front, based on a comparison of their coordinates
with the vertices of the triangle, which is implemented by a fast binary
search. When there is more than one point of the shapefile in any triangle, we
condense them into a single constraint, obtained by adding the relevant rows
of $H$. This way, we avoid over constraining the fire arrival time near the
perimeter, which should be avoided for the same reason as limiting the number
of constraints in mixed finite elements to avoid locking, cf.,
e.g.,~\cite{Brezzi-1991-MHF}.

\subsection{Numerical minimization of the residual}

\label{sec:nummin}

To solve (\ref{eq:constrained-min}) numerically, we use a multiscale descent
method similar to multigrid, combining line searches in the direction of
changes of the value of $T$ at a single point, and linear combinations of
point values as in~\cite{McCormick-1983-UMS}. We use bilinear coarse grid
functions with the coarse mesh step growing by a factor of 2. See
Fig.~\ref{fig:multigrid}(b) for an example of a coarse grid function with
distance between nodes 16 mesh steps on the original, finest level. We start
from an initial approximate solution that satisfies the constraint $HT=g$
exactly, and project all search directions on the subspace $Hu=0$, so that the
constraint remains satisfied throughout the iterations.

To find a reasonable\ initial approximation to the fire arrival time, we solve
the quadratic minimization problem%
\begin{equation}
I\left(  T\right)  =\frac{1}{2}\int\limits_{\Omega}\left\Vert \left(
-\triangle\right)  ^{\alpha/2}T\right\Vert ^{2}dxdy\rightarrow\min_{T}\text{
subject to (\ref{eq:perimeter-conditions}) and }\frac{\partial T}{\partial\nu
}=0, \label{eq:laplace}%
\end{equation}
where $\nu$ is the normal direction, $\triangle=\frac{\partial^{2}}{\partial x^{2}}+\frac{\partial^{2}%
}{\partial y^{2}}$ is the Laplace operator, and $\alpha>1$ is generally
non-integer. The reason for choosing $\alpha>1$ is that $\sqrt{I\left(
T\right)  }$ is the Sobolev $W^{\alpha,2}\left(  \Omega\right)  $ seminorm and
in 2D, the space $W^{\alpha,2}\left(  \Omega\right)  $ is embedded in
continuous functions if and only if $\alpha>1$. Consequently, $I\left(
T\right)  $ is not a bound on the value $T\left(  x,y\right)  $ at any
particular point, only averages over some area can be controlled. Numerically,
when $\alpha=1$, minimizing $I\left(  T\right)  $ with a point constraint,
such as an ignition point, results in $T$ taking the shape of a sharp funnel
at that point (Fig.~\ref{fig:initial}), which becomes thinner as the mesh is
refined. That would be definitely undesirable.

The discrete form of (\ref{eq:laplace}) is
\begin{equation}
\frac{1}{2}\left\langle ST,T\right\rangle -\left\langle f,T\right\rangle
\rightarrow\min_{T}\text{ subject to }HT=g,\label{eq:constr-discrete}%
\end{equation}
where $S=A^{\alpha}$ with $(-A)$ a discretization of the Laplace operator with
Neumann boundary conditions. To solve (\ref{eq:constr-discrete}), we first
find a feasible solution $u_{0}=H^{\prime}\left(  HH^{\prime}\right)  ^{-1}g$,
so that $Hu_{0}=g$, substitute $T=u_{0}+v$ to get
\[
\frac{1}{2}\left\langle S\left(  u_{0}+v\right)  ,u_{0}+v\right\rangle
-\left\langle f,u_{0}+v\right\rangle \rightarrow\min_{T}\text{ subject to
}Hv=0,
\]
and augmenting the cost fuction, we get that (\ref{eq:constr-discrete}) is
equivalent to%
\begin{equation}
\frac{1}{2}\left\langle SPv,Pv\right\rangle +\frac{\rho}{2}\left\langle
\left(  I-P\right)  v,v\right\rangle -\left\langle f_{0},v\right\rangle
\rightarrow\min_{T}\text{ subject to }Hv=0\text{,}\label{eq:constr-homog}%
\end{equation}
where $f_{0}=f-Su_{0}$, $P=I-H^{\prime}\left(  H^{\prime}H\right)  ^{-1}H$ is
the orthogonal projection on the nullspace of $H$, and $\rho>0$ is an arbitrary
regularization parameter. We solve the minimization problem
(\ref{eq:constr-homog}) approximately by preconditioned conjugate gradients
for the equivalent symmetric positive definite linear system%
\begin{equation}
P\left(  SPv-f_{0}\right)  +\rho\left(  I-P\right)  v=0.\label{eq:linear}%
\end{equation}
Since $S$ is discretization of the Neumann problem, the preconditioner
requires some care. Define $Z$ as the vector that generates the nullspace of $S$,
which consists of the discrete representation of constant functions, and
$P_{Z}=I-Z^{\prime}\left(  Z^{\prime}Z\right)  ^{-1}Z$ the orthogonal
projection on its complement. We use the preconditioner
\[
M:r\mapsto PP_{Z}S^{+}P_{Z}Pr,
\]
where $S^{+}$ is the inverse of $S$ on the complement of its nullspace, and
recover the solution by $T=u_{0}+Pv$. The method only requires access to
matrix-vector multiplications by $S$ and $S^{+}$, which are readily
implemented by cosine FFT. We only need to solve (\ref{eq:linear}) to low
accuracy to get a reasonable starting point for the nonlinear iterations, but
the satisfaction of the constraint $HT=g$ to rounding precision is important.%



\section{Assimilation of MODIS and VIIRS fire detections}

\label{sec:assim}

Data likelihood is the probability of a specific configuration of fire
detection and non-detection pixels given the state of the fire. The
probability of MODIS Active Fires detection in a particular sensor pixel as a
function of the fraction of the area actively burning and the maximum size of
contiguous area burning, was estimated in the validation
study~\cite{Schroeder-2008-VGM} using logistic regression. We consider the
fraction of the pixel burning and the maximum continuous area burning as a
proxy to the fire radiative heat flux in the pixel. The model state is encoded
as the fire arrival time at each grid point, and the heat flux can be then
computed from the burn model using the fuel properties. Substituting the heat
flux into the logistic curve yields a plausible probability of detection for a
period starting from the fire arrival time: the probability keeps almost
constant while the fire is fresh, and then diminishes.

However, the position uncertainty of the detection is significant, the allowed
$3\sigma$-error is listed in VIIRS specifications~\cite{Sei-2011-VAF-x} as
$1.5$km, and position errors of such magnitude are indeed occasionally
observed. Therefore, the probability of detection at the given coordinates of
the center of a sensor pixel in fact depends on the fire over a nearby area,
with the contributions of fire model cells weighted by $e^{-d^{2}/\sigma^{2}}%
$, where $d$ is the distance of the fire model cell and the nominal center of
the sensor pixel, because of the uncertainty where the sensor is actually
looking. Assuming that the position errors and the detection errors are
independent, we can estimate the contribution of a grid cell to the data
likelihood from a combination of the probabilities of detection at the nearby
satellite pixels.

Assimilation of data into the fire spread model can be then formulated as an
optimization problem to minimize its residual and to maximize the data
likelihood. See~\cite{Haley-2018-DLA} for further details.

\begin{figure}[ptb]
\centering
\includegraphics[width=\textwidth]{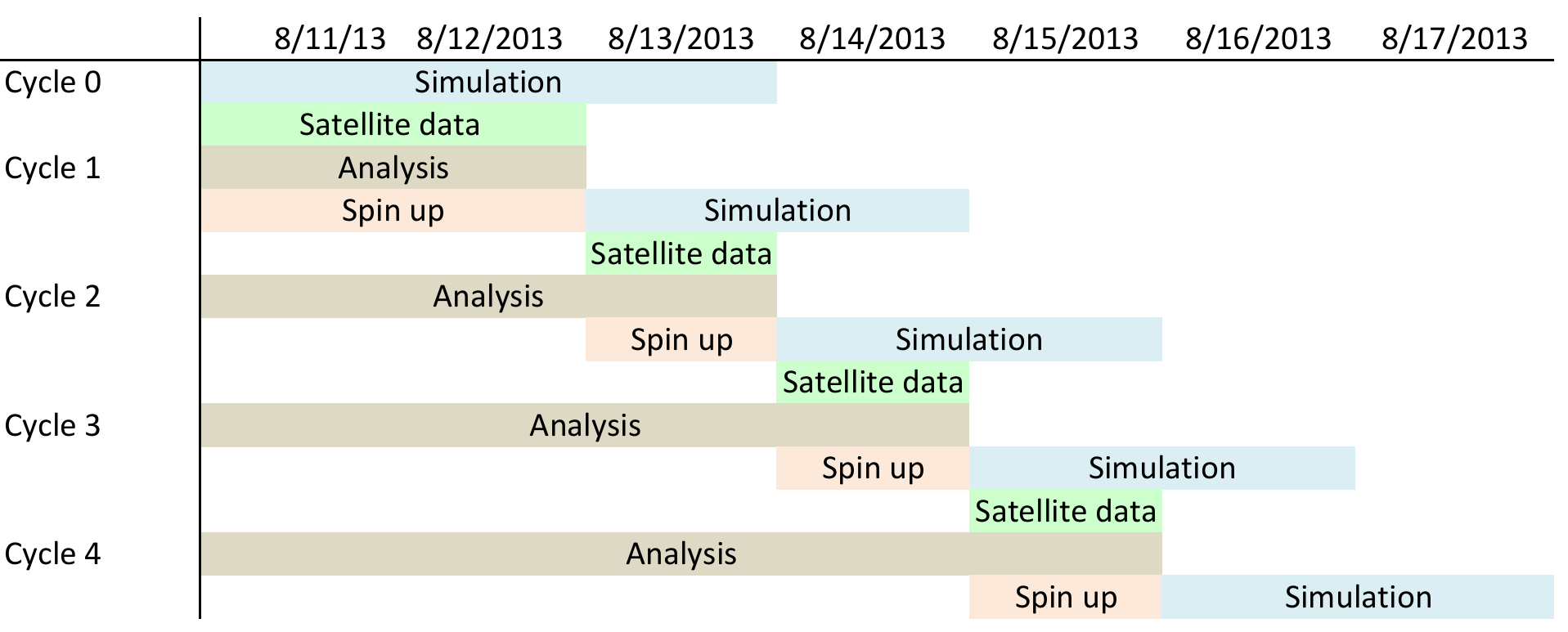}\caption{Data
assimilation cycling with atmosphere model spin up.
From~\cite{Mandel-2016-AMV}.}%
\label{fig:cycling}%
\end{figure}\begin{figure}[pt]
\centering
\includegraphics[scale=0.35]{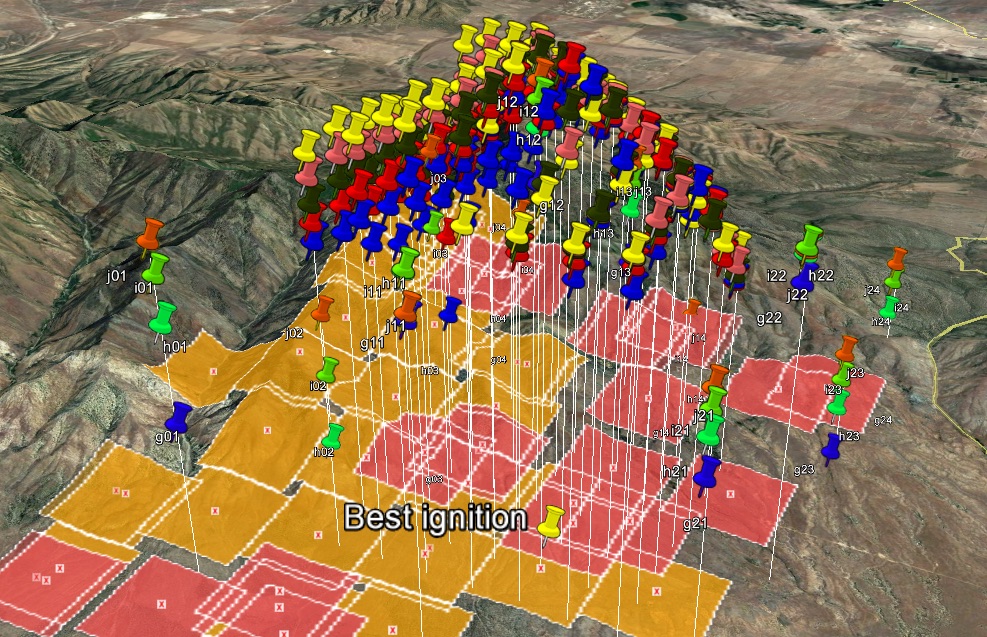}\caption{Estimation of the
most likely time and ignition point of a fire by evaluation of MODIS Active
Fire data likelihood. The color of the pushpin represents the time of ignition
and the height of the pushpin gives the likelihood of ignition at that
location.}%
\label{fig:pushpin}%
\end{figure}

Since the fire model is coupled with an atmosphere model, changing the state
of the fire alone makes the state of the coupled model inconsistent. To
recover a consistent state, we spin up the atmosphere model from an earlier
time, with the modified fire arrival time used instead of the fire arrival
time from the fire spread model (Fig.~\ref{fig:cycling}). This synthetic fire
forcing to the atmospheric model is used to drive atmospheric
model~\cite{Mandel-2012-APD} and enables establishing fire-induced circulation.

Varying the model state to maximize the data likelihood can also be used to
estimate the time and place of ignition as well as other model parameters. The
WRF-SFIRE \cite{Mandel-2011-CAF} model was run on a mesh of varying GPS
coordinates and times and the data likelihoods of the relevant Active Fire
detection data is evaluated, allowing the most likely place and time of the
fire's ignition to be determined. Fig.~\ref{fig:pushpin} shows a visualization
of the likelihoods of Active Fire detection data for several hundred ignition
points at various times. Work is in progress so that an automated process of
determining the most likely time and place of ignition can be initiated from
collection of satellite data indicating a wildfire has started in a particular
geographic region of interest.

%



\section{Computational experiments}

\begin{figure}[ptb]
   \centering
\begin{tabular}{cc}
\includegraphics[width=5cm]{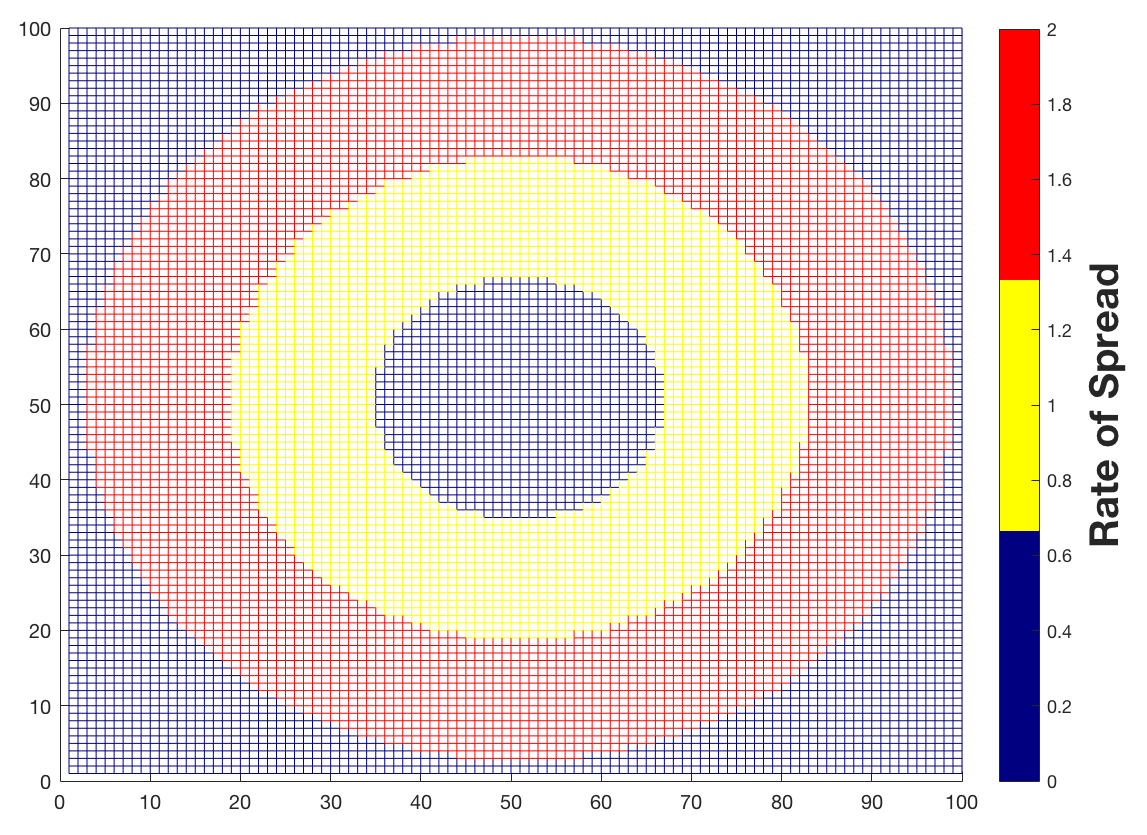}&
\includegraphics[width=7cm]{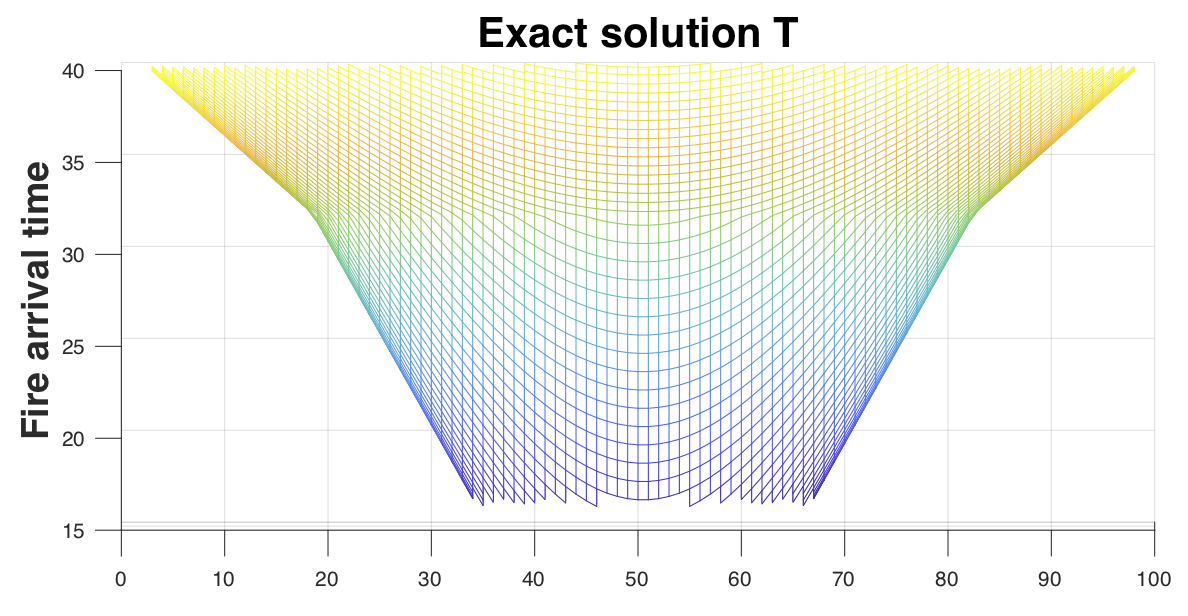}\\
(a) & (b)
\end{tabular}
    \caption{(a) Initial approximation of the fire arrival time $T$ in the two concentric circles perimeter case using different values of $\alpha$. (b) Exact solution $T$ for the concentric circles problem.}
    \label{fig:conc}
\end{figure}


\begin{figure}[ptb]
   \centering
\begin{tabular}{cc}
\includegraphics[width=6cm]{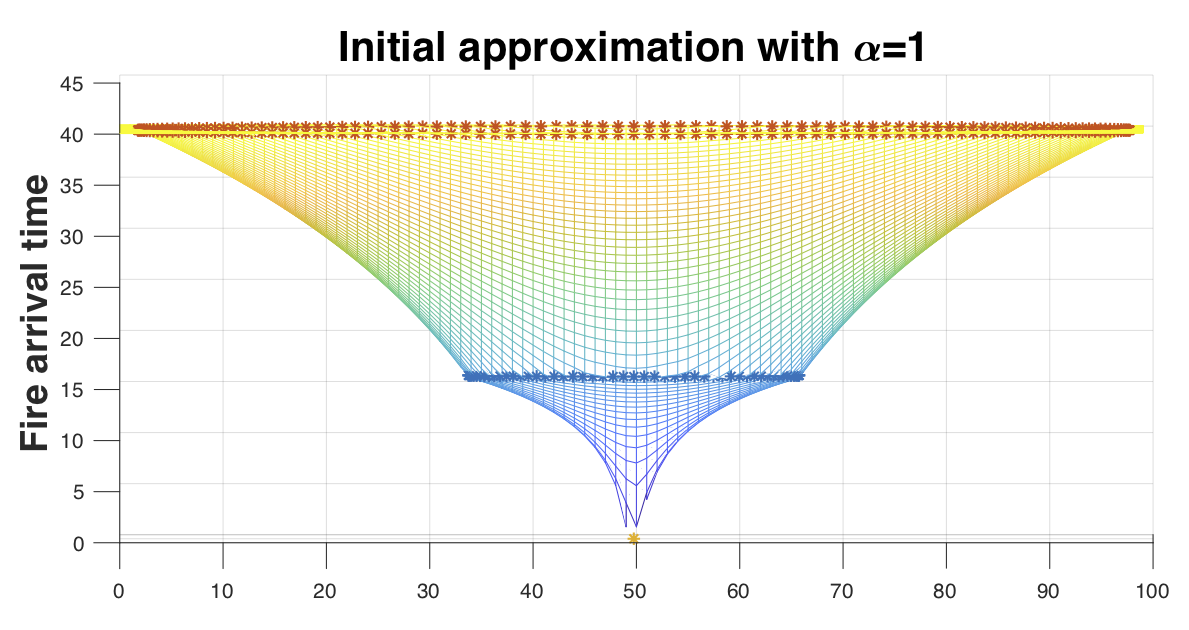}&
\includegraphics[width=6cm]{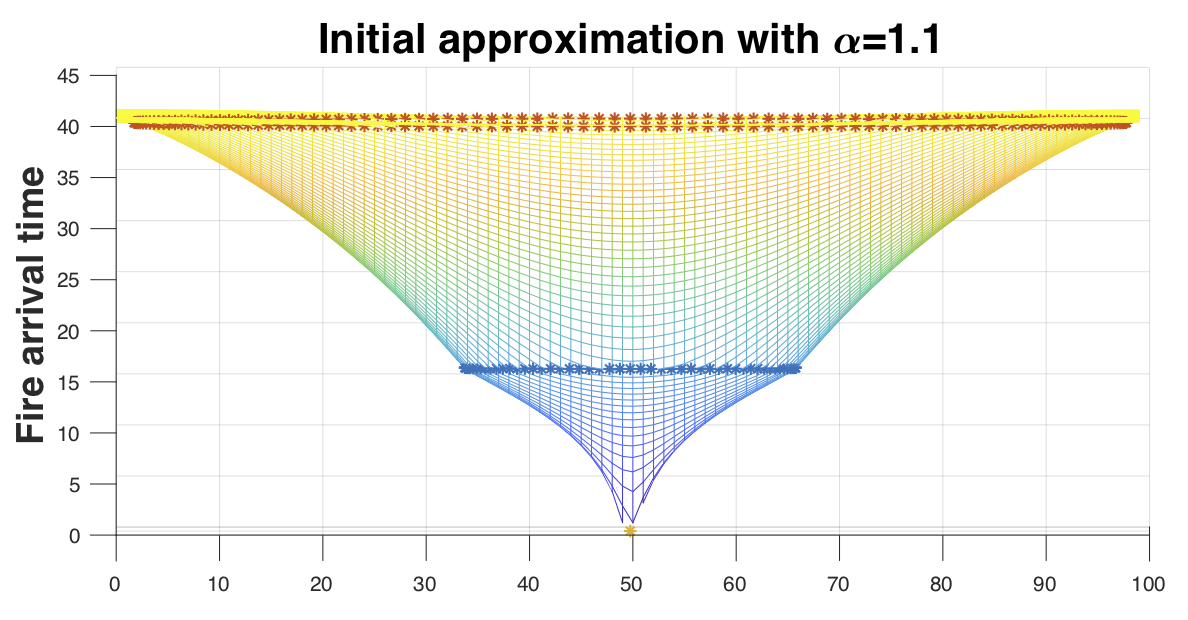}\\
(a) & (b) \\
\includegraphics[width=6cm]{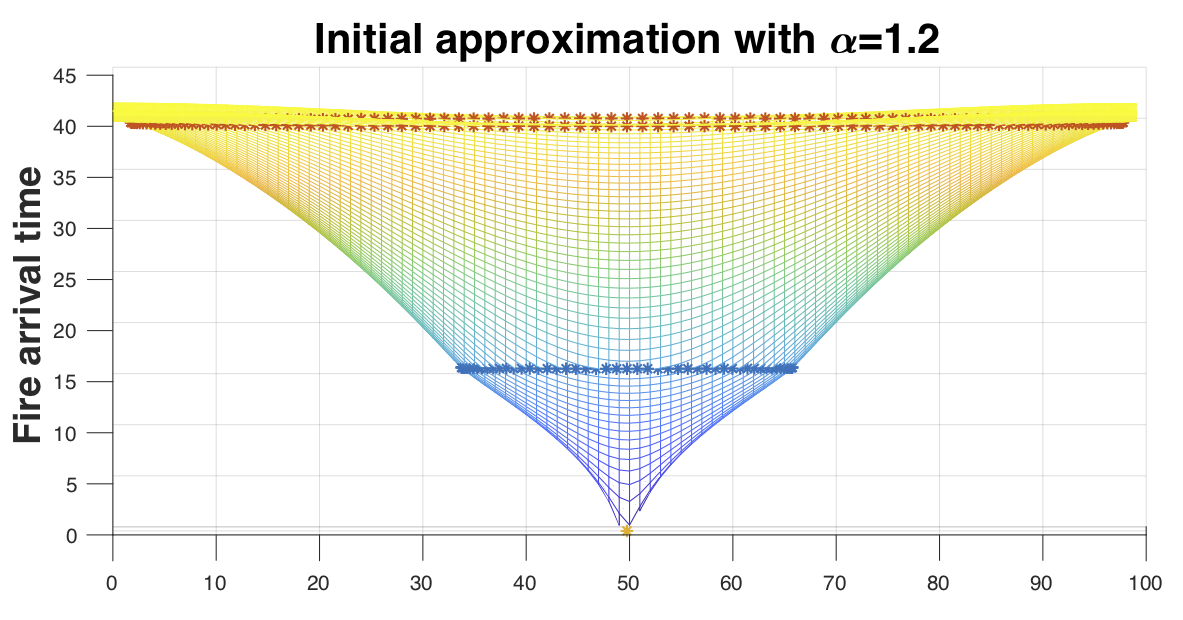}&
\includegraphics[width=6cm]{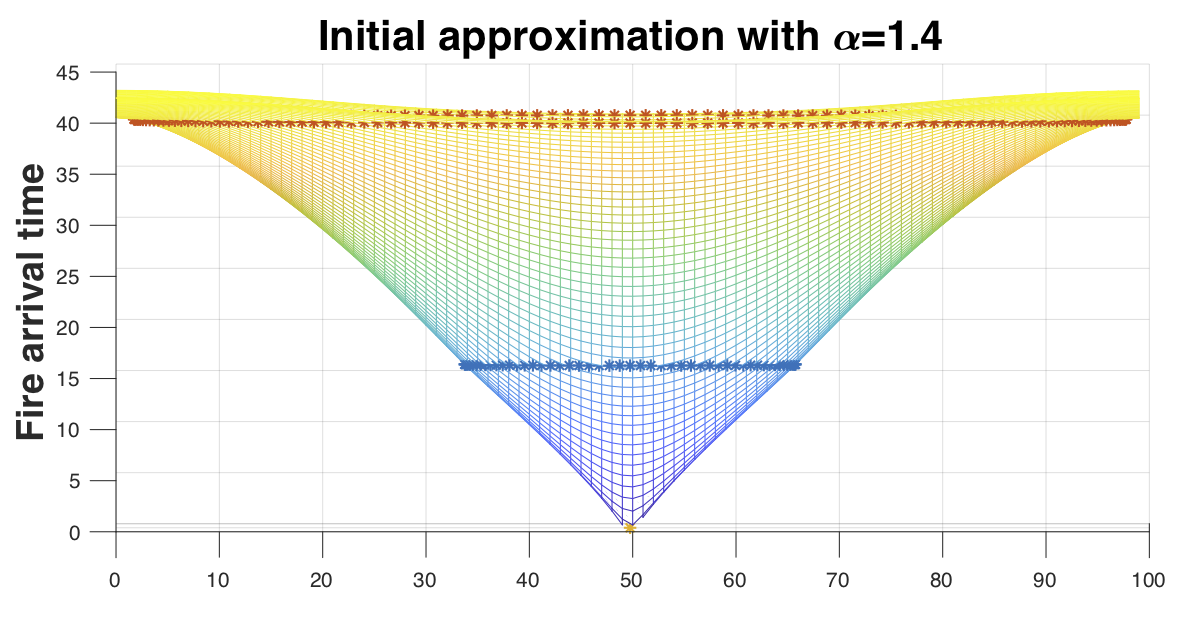}\\
(c) & (d)
\end{tabular}
    \caption{Initial approximation of the fire arrival time $T$ in the two concentric circles perimeter case using different values of $\alpha$.}
    \label{fig:initial}
\end{figure}

\begin{figure}[ptb]
   \centering
\begin{tabular}{cc}
\includegraphics[width=5.8cm]{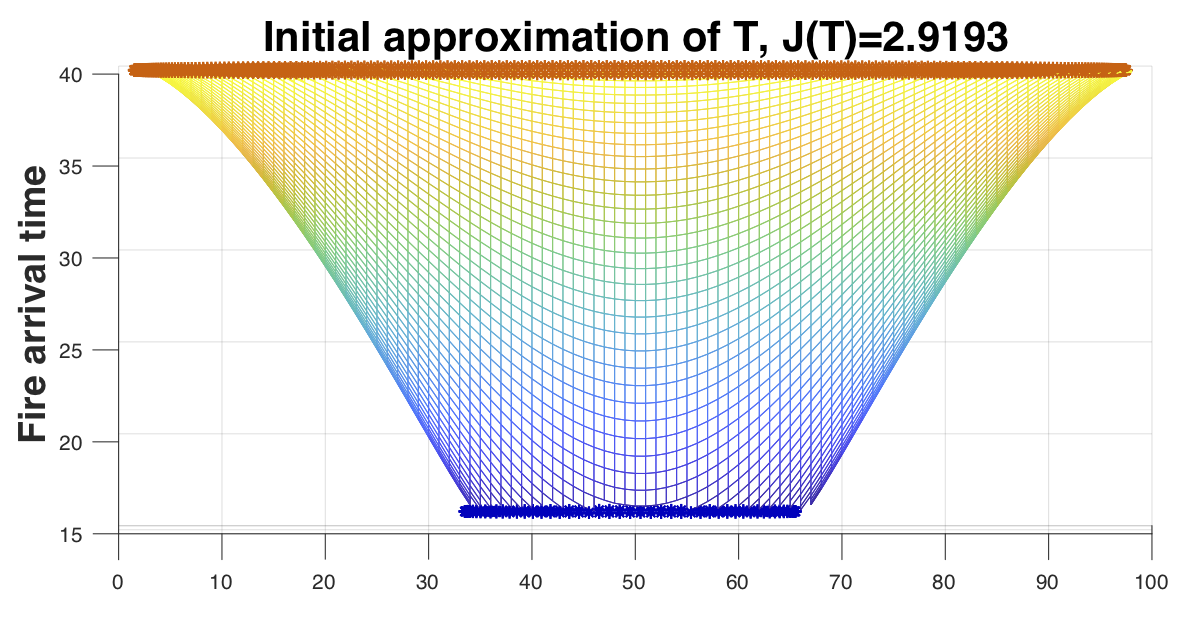}&
\includegraphics[width=6.2cm]{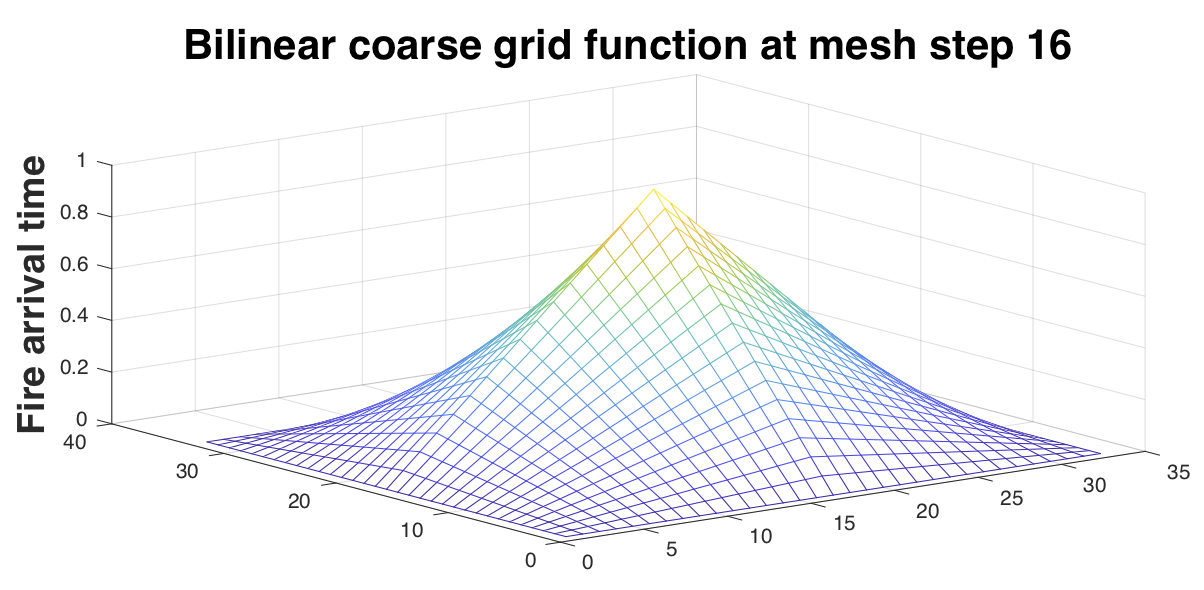}\\
(a) & (b) \\
\includegraphics[width=5.8cm]{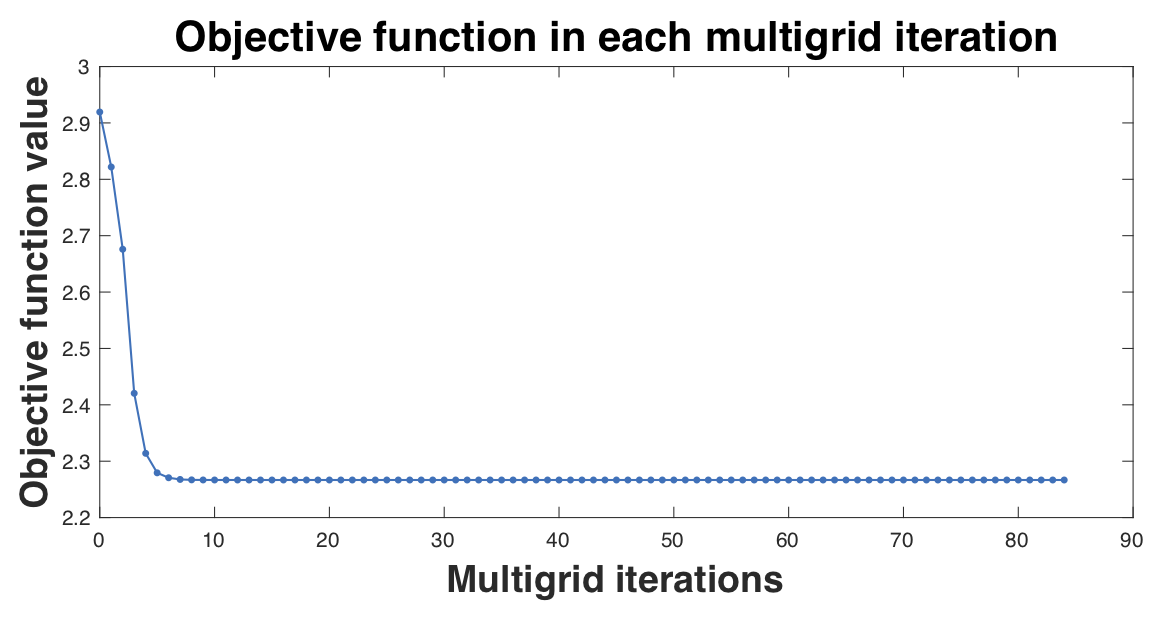}&
\includegraphics[width=6cm]{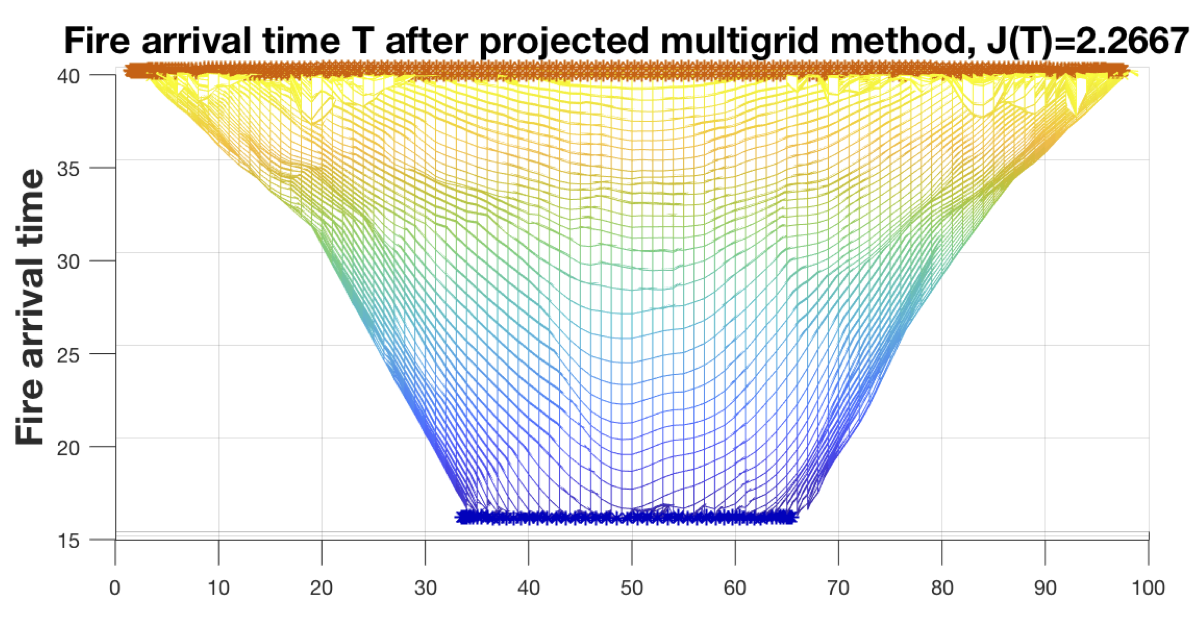}\\
(c) & (d)
\end{tabular}
    \caption{(a) Initial approximation from the first perimeter at $T_{1}=16$ to the second perimeter at $T_{2}=40$ obtained with $\alpha=1.4$. (b) Example of a bilinear coarse grid function at mesh step 16. (b) Values of the objective function after each line search iteration of the multigrid experiment. (c) Result of the fire arrival time interpolation after 4 cycles of multigrid
experiment.}
    \label{fig:multigrid}
\end{figure}


The optimization problem was tested on an idealized case using concentric
circles as perimeters in a mesh with $100\times100$ nodes. The fire spreads
equally in all directions from the center of the mesh. The propagation is
set at different rates of spread in different sections (Fig.~\ref{fig:conc}(a)).
We also set the fire arrival time at the ignition point and compute the fire
arrival time on the two perimeters from the given rate of spread, so in this
case there exists an exact solution  (Fig.~\ref{fig:conc}(b)).


The constraint matrix was constructed by the method described in
Sect.~\ref{sec:H}. The initial approximation of the fire arrival time was then
found by solving the quadratic minimization problem described in
Sect.~\ref{sec:nummin} with $\alpha=1.4$. Fig.~\ref{fig:initial} shows the
initial approximation of the fire arrival time imposed by the ignition point
and the two concentric circles in our particular case and using different
values of $\alpha$ from 1 to 1.4. One can see how the unrealistic sharp funnel
at the ignition point for $\alpha=1$ disappears with the increasing value of
$\alpha$.

Then, we run the multigrid method proposed in Sect.~\ref{sec:nummin}. The
coarsening was done by the ratio of 2. The number of sweeps was linearly
increasing with the level. On the coarsest level, the mesh step was 32 and the
sweep was done once, the mesh step on the second level was 16 and the sweep
was repeated twice, until resolution 1 on the original, finest grid, and sweep
repeated 6 times.

Fig.~\ref{fig:multigrid}c shows the decrease in the cost function with the
number of line searches on any level. One can observe that the cost function
decreased more in the first cycle and at the beginning of iterations on each level.

The final result after 4 cycles of 6 different resolutions (from 32 to 1
decreasing by powers of two) is shown in Fig.~\ref{fig:multigrid}(d),
which is close to the exact solution.


\section{Conclusions}

We have presented a new method for fitting data by an approximate solution of
a fire spread model. The method was illustrated on an idealized example.
Application to a real problem are forthcoming.

\section{Acknowledgments}

This research was partially supported by grants NSF ICER-1664175 and NASA
NNX13AH59G, and MINECO-Spain under contract TIN2014-53234-C2-1-R.
High-performance computing support at CHPC at the University of Utah and
Cheyenne (doi:10.5065/D6RX99HX) at NCAR CISL, sponsored by the NSF, are
gratefully acknowledged.

\label{sect:bib}
\bibliographystyle{splncs04}
\bibliography{../../references/geo,../../references/other,../../references/bddc,../../references/slides}



\end{document}